\documentclass{svjour3}                     % onecolumn (standard format)
\smartqed  % flush right qed marks, e.g. at end of proof
\usepackage{graphicx}
\journalname{General Relativity and Gravitation}
\begin{document}
\title{Gravitational wave background from neutron star phase transition}


\author{Jos\'e Carlos N. de Araujo         \and
        Guilherme F. Marranghello.
}


\institute{J C N de Araujo \at
              Instituto Espacial de Pesquisas Espaciais - Divis\~ao de Astrof\'{\i}sica \\
              Tel.: +55 12 39457223 \\
              Fax: +55 12 39456811\\
              \email{jcarlos@das.inpe.br}            \\
%                        \and
           G F Marranghello \at
              Universidade Federal do Pampa
}

\date{Received: date / Accepted: date}

\maketitle

\begin{abstract}
We study the generation of a stochastic gravitational wave (GW) background produced by a
population of neutron stars (NSs) which go over a hadron-quark phase transition in its
inner shells. We obtain, for example, that the NS phase transition, in cold dark matter
scenarios, could generate a stochastic GW background with a maximum amplitude of $h_{\rm
BG} \sim 10^{-24}$, in the frequency band $\nu_{\rm{obs}} \simeq 20-2000\, {\rm Hz}$ for
stars forming at redshifts of up to $z\simeq 20.$ We study the possibility of detection of
this isotropic GW background by correlating signals of a pair of Advanced LIGO
observatories. 
%\PACS{PACS code1 \and PACS code2 \and more} % \subclass{MSC code1 \and MSC code2 \and more}
\end{abstract}

\section{Introduction}

The first scientific runs on the gravitational wave (GW) observatories have already been done and the first results have been obtained. In particular, we take the american interferometer LIGO that has already put some limits on the properties of neutron star (NS) ellipticities \cite{benjamin,haskell}, for example. Putting some lower limits on such properties, like ellipticity, LIGO, among others like Virgo, GEO600 and TAMA, have contributed on the better understanding of NS properties.

NSs are excellent candidates to the first detection of GWs, with black holes (BHs), due to its great amount of matter constrained into a very small radius. NSs are astrophysical objects on which a great amount of quadrupole moment variation can be generated by a mountain on the star, a crust quake, a core quake, a glitch, rotation or collapse. All these events would generate and radiate GWs.

In this work, we address special attention to the GWs generated after a micro-collapse of the NS. Such event occurs, in a catastrophic way, when the NSs suffer a phase transition in its interior. This phase transition, in this work, is assumed by the releasing of quarks from the interior of hadrons. When the critical density is reached, quarks deconfine. As the quark matter equation of state is softer then the hadronic one, the NS shrinks.

It is widely accepted that NSs are born with fast rotating angular velocities. Due to magnetic torques, however, the NS periods could well be spun down. This spin-down causes a reduction in the centrifuge force and, consequently, the central energy density of the NSs increases. Those stars, born with densities close to that of the quark deconfinement, may undergo a phase transition forming a strange quark matter core. It is worth stressing, however, that it is not a common sense that such a transition really takes place.

As a consequence of such a putative phase transition, the stars could suffer a collapse
which could excite mechanical oscillations. The great amount of energy generated in this
process, $\Delta E\sim 10^{53}\,ergs$ ($\sim 0.1\, M_{\odot}c^{2}$) \cite{marr1}, could be
dissipated, at least partially, in the form of GWs, which are studied
in the present work.

The energy released to excite these mechanical oscillations can be driven into many different modes, such as the fundamental f-mode, the pressure p-mode, which are the overtones of the fundamental mode, g-modes, excited by a sharp difference in the density distribution, the w-mode, which is a pure relativity mode. The energy is also driven to heat the star, releasing neutrinos or ejecting the crust material \cite{miniutti}.

The number of NSs that undergo a phase transition was estimated to be about $10^{-6}
events\, / year\, / galaxy$ \cite{marr2}. However, this number depends on the properties of
the phase transition and can increase (drop) to more (less) significant values. We
investigate, in this work, the formation of a stochastic background of GWs originated by
such event in scenarios of cosmological structure formation.

We have improved the calculation made by Sigl\cite{sigl} by using a realistic equation of
state for describing NS matter, previously studied with politropic equations of state. For the
rate of events, we have used the results obtained by Marranghello {\it et.
al.}\cite{marr2}. Finally, the amount of energy released in the GW mode is
also estimated by realistic calculations made in ref.\cite{marranga}. Moreover, we adopted
in the present paper the history of star formation derived by \cite{springel}, who employed
hydrodynamic simulations of structure formation in a $\Lambda$ cold dark matter
($\Lambda$CDM) cosmology. A larger set of variables, such as the initial mass function and
progenitor masses, is also used for a richer analysis. Even though the results do not show
great differences, once the physics included is more realistic, the model becomes more
reliable.

Springel \& Hernquist \cite{springel} obtain the history of star formation from hydrodynamic simulations of
structure formation in ${\rm \Lambda CDM}$ cosmology. They study the history of cosmic star
formation from the ``dark ages", at redshift  ${\rm z} \sim 20$, to the present. They take
into account besides gravity and ordinary hydrodynamics, radiative heating and cooling of
gas, supernova feedbacks, and galactic winds. Their paper improves previous studies which
consider either semi-analytical models or numerical simulations.

It is worth mentioning that the story of star formation they obtain is consistent with
observations. It is important to bear in mind, however, that nowadays observations give
information from at most a redshift around ${\rm z} \sim 5$. In the future, however, with
the {\it Next Generation Space Telescope} (NGST) it will be possible to trace the cosmic
star formation rate out to ${\rm z} \geq 20$ (see, e.g., \cite{mackey}).

Besides the reliable history of star formation by Springel \& Hernquist, we consider the
role of the parameters $\alpha_{i=1,2}$, which gives the fraction of the progenitor mass which forms
the remnant NSs. We consider that the remnant mass is given as a function of the progenitor mass $M_r=\alpha_1m+\alpha_2$. Recall that a given initial mass function (IMF) refers to the distribution function of the stellar progenitor mass, and also to the masses of the remnant compact objects left as a result of the stellar evolution. In section 3 we show in detail why $M_r$ is written as above.

In the present study we have adopted a stellar generation with a Salpeter IMF, which is
consistent with Springel \& Hernquist, since they show that population II stars could have
been formed at high redshift too. We then discuss what conclusions would be drawn whether
(or not) the stochastic background studied here is detected by the forthcoming GW
observatories such as LIGO and VIRGO.

The paper is organized as follows. Section 2 deals with the NS equations of state adopted in our study;
in Section 3 we consider the IMF and the NS masses; Section 4 deals with the GW production;
in Section 5 we present the numerical results and discussions; in Section 6 the detectability
of the background of GWs are considered and finally in Section 7 we present the
conclusions.

\section{Neutron stars and the nuclear matter equation of state}

The global properties of mass and radius of compact stars are directly related to its equation of state.
In the present study we do not adopt a polytropic equation of state, since this could result in a poor modeling.
In particular, we consider a model developed by \cite{taurines} for a new class of parameterized field-theoretical model described by a Lagrangian density where the whole baryon octet is coupled to scalar and vector fields through parameterized coupling constants.

The main characteristic of this model is to adjust these parameterized coupling constants to fit both the nuclear matter properties and the neutron stars observation data. Varying this parametrization, we are also able to fit the properties of some other field-theoretical models like the Walecka model and the Zimanyi-Moskowski model and compare our results to those of other relevant models in the literature as described by \cite{taurines}. We are also able to reproduce calculations with higher and lower values for neutron star masses and radii, compression modulus and effective baryon masses.

Instead of performing the following calculations with a wide range of equations of state we preferred to describe our work choosing a representative equation of state whose coupling constants were chosen to be those that fit both the nuclear and stellar properties in an acceptable manner. It is important to notice that other parameterizations of this model could result in quite similar results, harder detection of the background or unreliable results due to the not well fitted characteristics of nuclear matter.

The nuclear matter Lagrangian density is described by
\begin{eqnarray}
{\cal L}    &=& \sum\limits_{B}   \bar{\psi}_{B}\left( i\gamma_\mu
(\partial^\mu- g^\star_{\omega B} \omega^{\mu}) -
(M_B-g^\star_{\sigma B} \sigma) - [\frac12 g^\star_{\varrho B}
\mbox{\boldmath$\tau$} \cdot \mbox{\boldmath$\varrho$}^\mu]
\right)\psi_B   \nonumber \\ && +\frac12(\partial_\mu \sigma
\partial^\mu \sigma   - {m_\sigma^2} \sigma^2)  - \frac14
\omega_{\mu \nu}  \omega^{\mu \nu}   + \frac12 {m_\omega^2}
 \omega_\mu \omega^\mu
\nonumber \\ && -   \frac14 \mbox{\boldmath$\varrho$}_{\mu \nu}
\cdot \mbox{\boldmath$\varrho$}^{\mu \nu} +  \frac12m_\varrho^2
\mbox{\boldmath$\varrho$}_\mu  \cdot  \mbox{\boldmath$\varrho$}^\mu
+\sum\limits_{l}   \bar{\psi}_{l} [i \gamma_\mu \partial^\mu   -
M_l] \psi_l  \,\, , \label{eqdl}
\end{eqnarray}
where the baryon field $\psi_B$ is summed over the whole baryon
octet and coupled to the scalar and vector fields $\sigma$, $\omega$
and $\varrho$ through the parameterized coupling constants
$g^\star_{\sigma B}$, $g^\star_{\omega B}$, $g^\star_{\varrho B}$.
The free lepton fields $\psi_\lambda$ contributes to the electrical
equilibrium in the NS matter. The masses of the baryons, mesons and
leptons are represented by $M_B$, $m_{\sigma,\omega,\varrho}$ and
$m_\lambda$, respectively. The free lepton fields contributes to the
electrical equilibrium in the NS matter.

Using a parametrization for the baryon-meson coupling constants,
\begin{equation}
g^\star_\sigma{\bar\psi}\sigma\psi=\frac{g_\sigma\sigma}{\left(1+
    \frac{g_\sigma\sigma}{\lambda M}\right)^\lambda}{\bar\psi}\psi
\end{equation}
this model describes a wide range of NS parameters, such as a maximum mass ranging from (very low values)
$M=0.66M_\odot$ up to $M=2.77M_\odot$, and the corresponding radii that vary in the range
of $8<R<13\, km$. Each pair in the mass-radius relation is associated with a different
parametrization of the equation of state (see \cite{taurines} for further details). Obviously, each value
of such parametrization represents different values of nuclear matter properties.

\begin{figure}
\includegraphics[width=0.5\textwidth,angle=-90]{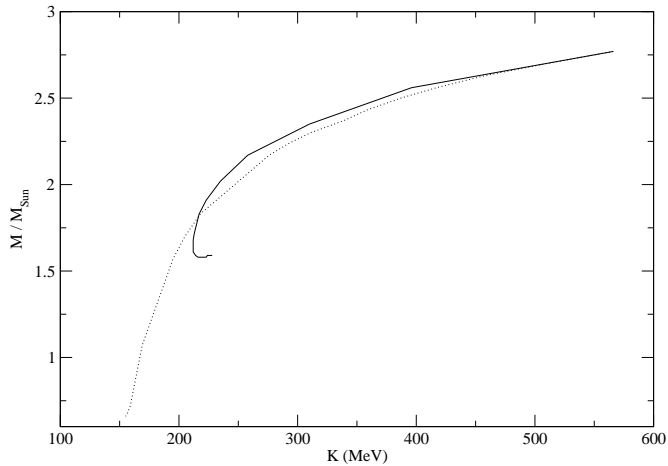} \caption{\label{1} The NS maximum
masses versus the compression modulus, for different parametrization in the Taurines {\it
et.al.} models. The solid line represents a parametrization in the scalar mesons while the
dotted line describes a parametrization in both scalar and vector contributions of the
nuclear force.}
\end{figure}

The study of nuclear matter properties, however, is still a very open problem. A relevant parameter
is the compression modulus, which defines the curvature of the equation of state or, in a few words, is
related to the capability of matter to be compressed. One can see in Figure \ref{1} that
the maximum mass of NSs is obtained for large values of the compression modulus. The
definition of $K$ reads

\begin{equation}
K=9\left[\rho^2\frac{d^2(\epsilon/\rho)}{d\rho^2}\right]_{\rho=\rho_0}
\end{equation}
where $\epsilon$ is the energy density, $\rho$ is the baryon density and $\rho_0$ the
baryon saturation density. Actually, the acceptable values for the compression modulus are
constrained to be in range $200<K<300MeV$ \cite{blaizot,myers}.

An important constraint to the equation of state is to describe a family of NSs with a maximum mass
greater than the value obtained by the most massive known pulsar. The PSR J0751+1807 has a
mass $M= 2.1\pm 0.2M_\odot$ \cite{nice}. This condition has to be fulfilled by the set of
parameters chosen to describe the NS model. This first condition already constrains the
value of compression modulus to $K\ge 220 MeV$ when this model is considered. Smaller
values of K would lead to masses smaller than $1.9M_\odot$ disagreeing with the measures of PSR J0751+1807 mass.

It is worth mentioning that quite recently \cite{freire} reported the discovery of eight new millisecond pulsars in NGC 6440 and NGC 6441. In particular, they could determined the mass of the pulsar J1748-2021B (NGC 6440B), namely $M= 2.74\pm 0.21M_\odot$. If the mass of  this pulsar is really high, this would mean that the value of the compression modulus would be around $500 MeV$. In other words, our model could describe, in an extreme limit of coupling constants, hybrid stars with that value of gravitational mass. However, it would lead to values of nuclear matter properties that are not in complete agreement with experiments.

Another scenario to be discussed concerns the strange quark star. The model used in this
analysis is the MIT bag model \cite{mit}. The MIT bag model describes, as its own name
suggests, that baryons are composed by three quarks confined inside a bag. The effects of
pressure difference in the interior and exterior regions of the bag are summarized in the
bag constant.

%The thermodynamical potential of quarks in the MIT bag model is described by

%\begin{eqnarray}
%\Omega & = & \sum_{q=u,d,s}\frac{-1}{4\pi^2}\left[\mu_q {k_F}_q\left(\mu_q^2-\frac 52 m_q^2\right)+\frac 32 m_q^4 \ln\left(\frac{\mu_q+{k_F}_q}{m_q}\right)\right] \nonumber \\
%& + & \frac{2\alpha_c}{4\pi^3}\left\{3\left[\mu_q {k_F}_q-m_q^2-m_q^2 \ln\left(\frac{\mu_q+{k_F}_q}{m_q}\right)\right]^2 -2{k_F}_q^4 - 3m_q^4 \ln^2\left(\frac{m_q}{\mu_q}\right) \nonumber \\
%& + &  6 \ln\left(\frac{\rho_r}{\mu_q}\right)\left[\mu_q {k_F}_q m_q^2-m_q^4 \ln\left(\frac{\mu_q+{k_F}_q}{m_q}\right)\right]\} + B
%\end{eqnarray}

%\noindent where $q=u,d,s$ are the three quark flavors up, down and strange,
%$\mu_q$ is the chemical potential, $m_q$ represents the quark
%masses, $k_q$ is the Fermi momentum of the particles, $B$ is the bag
%constant and the terms multiplied by $\alpha_c$ represents the
%contribution of the first order perturbation in the strong
%interaction.

Even with considerable advances in the study of QCD (quantum chromo-dynamics) on the
lattice, from where we expect the most reliable theoretical results and the consequent
advances in collision experiments, the bag constant still presents a very wide range of
possible values. In the same way, the perturbative constant, $\alpha_c$, the one that
represents the first order correction to the strong interaction forces between quarks, also
represents an open issue.

Once the transition to quark-gluon matter occurs, the weak interaction processes for the
quarks u, d and s
\begin{equation}
u + s \rightarrow d + u
\end{equation}
and
\begin{equation}
d + u \rightarrow u + s
\end{equation}
will take place, and a rapid transition occurs with a consequent gravitational
micro-collapse. As a result, a huge amount of energy is dissipated, some in the form of GWs
produced by quasi-normal modes excitation. The amount of energy released during the transition is calculated to be around $10^{53}erg$ according to ref.\cite{marr1}. Only a small part of this energy is released in the form of GWs \cite{marranga}.

\section{Initial mass function and the neutron star masses}

The calculation of the GW background from NS phase transition requires the knowledge of the distribution function of stellar masses, the so called stellar initial mass function (IMF), $\phi(m)$. Here the Salpeter IMF is adopted, namely

\begin{equation}
\phi(m) = A m^{-(1+x)},
\end{equation}

\noindent where $A$ is the normalization constant and $x=1.30$ (our fiducial value). The
normalization of the IMF is obtained through the relation

\begin{equation}
\int_{m_{\rm l}}^{m_{\rm u}} m\phi(m)dm = 1,
\end{equation}

\noindent where we consider $m_{\rm l} = 0.1\, {\rm M}_{\odot}$ and $m_{\rm u}=125\, {\rm
M}_{\odot}$. For further details we refer the reader to, e.g., \cite{araujo2002}.

It is worth mentioning that concerning the star formation at high redshift, the IMF could
be biased toward high-mass stars, when compared to the solar neighborhood IMF, as a result
of the absence of metals \cite{bromm99,bromm02}.

On the other hand, in the study by Springel and Hernquist, besides a high-mass star
formation, it is shown that the population II stars, whose IMF could well be of the
Salpeter's type, could start forming around redshift 20 or higher.

In the present study we consider this population II studied by these authors. Then, for the standard IMF, the mass fraction of NSs produced as remnants of the stellar evolution is

\begin{equation}
f_{\rm NS} = \int_{\rm m_{min}}^{m_{\rm u}}M_{\rm r}\phi(m)dm,
\end{equation}

\noindent where $m_{\rm min}$ is the minimum stellar mass capable of producing a NS at the
end of its life, and $M_{\rm r}$ is the mass of the remnant NS. Stellar evolution
calculations show that the minimal progenitor mass to form NSs is $m_{\rm min}=8 {\rm
M}_{\odot}$, while the maximum progenitor mass is $m_{\rm max}= 25 - 40 {\rm M}_{\odot}$
(see, e.g., \cite{timmes}).

We do not intend to discuss in the present paper stellar evolution scenarios related to the formation of NSs. Many works can be found in the literature concerning the study of stellar evolution, supernova and NS formation\cite{janka}. However, we need to formulate a relation between the progenitor star mass $m$ and the remnant NS mass $M_{\rm  r}$. For the remnant, $M_{\rm  r}$, we take

$$M_{\rm r}=\alpha_1\, m+\alpha_2 \, ,$$

\noindent where $\alpha_1$ and $\alpha_2$ are constants; the first one is dimensionless and the second one is given in solar masses. As the results for stellar evolution are not yet fully determined, we studied this parametrization in three different scenarios for the values of $\alpha_1$ and $\alpha_2$ which would represent the distribution of NS mass as function of the progenitor mass (see Table 1 and the following sections). For example, considering NSs formed from progenitors with $m=20M_\odot$ we would have remnants with masses $M_r=$ 1.375$M_\odot$, 1.5$M_\odot$ and 2.7$M_\odot$. These parameters describe sharper or softer distribution of NS masses around the standard value of mass $M=1.4M_\odot$ . Note that we are considering in the end that $M_{\rm r}$ is independent of the redshift.

\begin{table}
\caption {The values of $\alpha_{i=1,2}$ which determines the fraction of the
  NS remant.}
\begin{center}
\begin{tabular}{lccc}
\hline
 $\alpha_1$             & 1/32 & 1/50 & 1/17 \\
 $\alpha_2$ \,($M_\odot$) & 3/4  & 1.1  & 0.53 \\
\hline
\end{tabular}
\end{center}
\end{table}

With these considerations at hand, the mass fraction of NSs reads up to $f_{\rm
NS}=10^{-2}$ for $x=1.30$, while the fraction of NSs that undergoes a phase transition
($f_{\rm NS}^{pt}$) can drop down to values $\ll 1$ (see, e.g., \cite{marr2}).

To assess the role of possible IMF variations in our results, other values of $x$ have also been considered. Besides the standard IMF, two others have been studied, namely, with
$x=0.3$ and $x=1.85$, which yield approximately ten times and one-tenth of mass fraction of NSs of the standard IMF, respectively.

In order to cover a wide number of parameters we present, in Table 2, the models considered in our study. In the first column appears the name of the model; in the second and third columns we present different combinations of the parameters $\alpha_{1}$ and $\alpha_{2}$, which imply in different ways to calculate the NS remnant mass for a given IMF; in the fourth column the mass range of the progenitor star; in the fifth column the NS remnant mass; in the sixth column the NS redshift formation, and finally in the seventh column the observed frequency, which are obtained via the empirical formulae given in reference \cite{benhar}.

These parameters are directly related to the number of NSs that goes over the phase transition. According to the NS models obtained by realistic equations of state, only a fraction of them develops a core composed by deconfined quarks and gluons. When we set the mass range of the progenitor mass, we are also setting the mass of the remnant object. Stars with higher mass reach central densities that are high enough to develop the quark core, while smaller stars do not. The available energy for GW emission generated in the collapse process is also implicit in the mass of the remnant star. The more massive is the star, greater is its core and greater is the energy difference between the neutron and the hybrid star. We have found that stars with baryonic mass $M_b=2M_\odot$ develops a core as large as $R=7km$ and can generate as much as $3\times10^{53}erg$ of energy, while stars with baryonic mass $M_b=1M_\odot$ develops a small core with $R=1km$ and has ten times less energy available from the transition. As we do not know for sure the amount of this energy that is driven in each mode and in order to simplify our calculations we have adopted the standard medium value of 0.01$M_\odot$ for the released energy.

\begin{table}
\caption {Models description: mass function constants, $\alpha_{1,2}$, progenitor, $m$, and
remnant, $M$, mass ranges, redshift, $z$, and the observed frequency, $\nu$.}
\begin{center}
\begin{tabular}{ccccccc}
\hline Model & $\alpha_1$ & $\alpha_2$ & $\Delta m (M_\odot)$ & $\Delta M (M_\odot)$ &
$\Delta z$ & $\Delta\nu (Hz)$ \\ \hline
A  & $1/32$ & $3/4 $ & 8.0-40.00  & 1.00-2.00 & 0-20 & 48-2000 \\
B  & $1/32$ & $3/4 $ & 14.4-16.32 & 1.20-1.26 & 0-50 & 31-1666 \\
C  & $1/32$ & $3/4 $ & 14.4-16.32 & 1.20-1.26 & 0-20 & 75-1666 \\
D  & $1/32$ & $3/4 $ & 20.8-22.72 & 1.40-1.46 & 0-20 & 65-1428 \\
E  & $1/50$ & $1.1 $ & 8.0-40.00  & 1.26-1.90 & 0-20 & 50-1587 \\
F  & $1/50$ & $1.1 $ & 15.0-18.00 & 1.40-1.46 & 0-20 & 65-1428 \\
G  & $1/17$ & $0.53$ & 8.0-25.00  & 1.00-2.00 & 0-20 & 48-2000 \\ \hline
\end{tabular}
\end{center}
\end{table}

\section{Gravitational wave production}

The GWs can be characterized by their dimensionless amplitude, $h$, and frequency, $\nu$.
The spectral energy density, the flux of GWs, received on Earth, $F_\nu$, in ${\rm erg}\;
{\rm cm}^{-2}{\rm s}^{-1}{\rm Hz}^{-1}$, is (see, e.g., \cite{douglass,hils}).

\begin{equation}
F_{\nu} = {c^{3}s_{\rm h}\omega_{\rm obs}^{2}\over 16{\rm \pi} G}, \label{fluxa}
\end{equation}

\noindent where $\omega_{\rm obs}=2{\rm \pi} \nu_{\rm{obs}}$, with $\nu_{\rm{obs}}$ the GW
frequency (Hz) observed on Earth, $c$ is the velocity of light, $G$ is the gravitational
constant and $\sqrt{s_{\rm h}}$ is the strain amplitude of the GW  in $\rm Hz^{-1/2}$.

The stochastic GW background produced by NS phase transition would have a spectral density
of the flux of GWs and strain amplitude also related to the above equation (\ref{fluxa}).
Therefore, in the above equation the strain amplitude takes into account the star formation
history occurring at the `first light', just after the `dark age' epoch. The strain
amplitude at a given frequency, at the present time, is a contribution of NSs with
different masses at different redshifts. Thus, the ensemble of NSs formed produces a
background whose characteristic strain amplitude at the present time is $\sqrt s_{\rm h}$.

On the other hand, the spectral density of the flux can be written as
\cite{ferraria,ferrarib}

\begin{equation}
F_{\nu}=\int_{z_{\rm {cf}}}^{z_{\rm {ci}}} \int_{m_{\rm {min}}}^{m_{\rm {u}}}
f_{\nu}(\nu_{\rm{obs}}) dR_{\rm NS}(m,z),
\end{equation}

\noindent where $f_{\nu}(\nu_{\rm{obs}})$ is the energy flux per unit of frequency (in
${\rm erg}\;{\rm cm}^{-2}{\rm Hz}^{-1}$) produced by the formation of a unique NS and
$dR_{\rm NS}$ is the differential rate of NSs formation.

The above equation takes into account the contribution of different masses that collapse to
form NSs occurring between redshifts $z_{\rm ci}$ and $z_{\rm cf}$ (beginning and end of
the star formation phase, respectively) that produce a signal at the same frequency
$\nu_{\rm{obs}}$. On the other hand, we can write $f_{\nu}(\nu_{\rm{obs}})$ \cite{carr} as

\begin{equation}
f_{\nu}(\nu_{\rm{obs}}) = {{\rm \pi} c^{3}\over 2G}h_{\rm NS}^{2},
\end{equation}

\noindent where $h_{\rm NS}$ is the dimensionless amplitude produced by the NS
micro-collapse of a given star with mass $m$ that generates at the present time a signal
with frequency $\nu_{\rm{obs}}$. Then, the resulting equation for the spectral density of
the flux is

\begin{equation}
F_{\nu} = {\pi c^{3}\over 2G} \int h_{\rm  NS}^{2}dR_{\rm NS}.
\end{equation}

\par\noindent From the above equations we obtain for the strain
amplitude

\begin{equation}
s_{\rm h} = {1 \over \nu_{\rm obs}^{2}}\int h_{\rm NS}^{2} dR_{\rm NS}.
\end{equation}

\par\noindent Thus, the dimensionless amplitude reads

\begin{equation}
h_{\rm BG}^{2} = {1 \over \nu_{\rm obs}}\int h_{\rm NS}^{2} dR_{\rm NS},
\end{equation}

\par\noindent (see \cite{araujo2000,araujo2005}).

The micro-collapse produces GWs at frequency $\nu$ of the NS f-mode, and dimensionless
amplitude given by \cite{ander1998}

\begin{equation} \label{h}
h_{NS} \simeq 1\times10^{-19}\left({\frac{E}{M_\odot c^2}}\right)^{1/2} \left(\frac{2
kHz}{\nu}\right)^{1/2}\left(\frac{1 Mpc}{d_{\rm L}}\right)
\end{equation}

\noindent where $E$ is the available pulsation energy, and $d_{\rm L}$ is the luminosity
distance to the source.

Recall that the energy available to excite the pulsating modes is directly related to the
equation of state adopted.

Note that, since we are considering cosmological sources we have to take into account the
redshift effect on the emission frequency, that is, a signal emitted at frequency $\nu_{\rm
e}$ at redshift $z$ is observed at frequency $\nu_{\rm{obs}}=\nu_{\rm e}(1+z)^{-1}$.

For the differential rate of NS we consider only those that undergo micro-collapse, namely
\begin{equation}
dR_{\rm NS} = \dot\rho_{\star}(z) {dV\over dz} f_{\rm NS}^{pt} \phi(m)dmdz,
\end{equation}
where $\dot\rho_{\star}(z)$ is the star formation rate (SFR) density (in ${\rm
M}_{\odot}\,{\rm yr}^{-1}\,{\rm Mpc}^{-3}$),  $dV$ is the comoving volume element, and
$f_{\rm NS}^{pt}$, as already mentioned, is the fraction of NSs which may undergo a phase
transition forming a strange quark matter core.

The comoving volume element is given by

\begin{equation}
dV = 4{\rm \pi}\bigg({c\over H_{0}}\bigg) r_{\rm z}^{2} {\mathcal{F}}(\Omega_{\rm
M},\Omega_{\Lambda},z) dz,
\end{equation}

\par\noindent with

\begin{equation}
{\mathcal{F}}(\Omega_{\rm M},\Omega_{\Lambda},z) \equiv {1\over \sqrt{(1+z)^2(1+\Omega_{\rm
M}z)-z(2+z)\Omega_{\Lambda}}},
\end{equation}

\noindent and the comoving distance, $r_{\rm z}$, is

\begin{equation}
r_{\rm z}={c\over H_{0}\sqrt{|\Omega_{\rm k}|}}S\bigg( \sqrt{|\Omega_{\rm
k}|}\int_{0}^{z}{dz'\over {\mathcal{F}}(\Omega_{\rm M},\Omega_{\Lambda},z')}\bigg),
\end{equation}

\par\noindent where

\begin{equation}
\Omega_{\rm M}=\Omega_{\rm DM}+\Omega_{\rm B} \qquad {\rm and} \qquad 1 = \Omega_{\rm
k}+\Omega_{\rm M}+\Omega_{\Lambda}
\end{equation}

\par\noindent are the usual density parameters for the
matter (M), i.e., dark matter (DM) plus baryonic matter (B), curvature (k) and cosmological
constant ($\Lambda$). The function S is given by

\begin{eqnarray}
  S(x) = \cases { \sin x  &if  closed, \cr x &if  flat, \cr \sinh x &if  open.}
\end{eqnarray}

\noindent The comoving distance is related to the luminosity distance by

\begin{equation}
d_{\rm L} = r_{\rm z} (1+z).
\end{equation}

The set of equations presented above can be used to find the dimensionless amplitude of
the GW background produced from a population of NSs which goes over a hadron-quark phase
transition in its inner shells as a function of the SFR density, and related to the `first
light' epoch.

For the SFR density, we adopt the one derived by Springel and Hernquist (see,
\cite{springel} for details), namely

\begin{equation}
\label{eqnfit} \dot\rho_\star(z)= \dot\rho_m\, \frac{\beta\exp\left[\Delta(z-z_m)\right]}
{\beta-\Delta+\Delta\exp\left[\beta(z-z_m)\right]},
\end{equation}

\noindent where $\Delta= 3/5$, $\beta=14/15$, $z_m=5.4$ marks a break redshift, and
$\dot\rho_m= 0.15\,{\rm M}_\odot{\rm yr}^{-1}{\rm Mpc}^{-3}$ fixes the overall
normalization.

It is worth mentioning that these authors employed hydrodynamic simulations of structure
formation in a $\Lambda$CDM cosmology with the following parameters: $\Omega_{\rm M}=0.3$,
$\Omega_{\Lambda}=0.7$, Hubble constant $H_{0}=100\; h\; {\rm km\;s^{-1}\;Mpc^{-1}}$ with
$h=0.7$, $\Omega_{\rm B}=0.04$, and a scale-invariant primordial power spectrum with index
$n=1$, normalized to the abundance of rich galaxy clusters at present day
($\sigma_{8}=0.9$).

Another relevant physical quantity associated with the GW background is
the closure energy density per logarithmic frequency span, which is given by

\begin{equation}
\Omega_{\rm GW} = {1\over \rho_{\rm c}} {d\rho_{\rm GW}\over d\log \nu_{\rm{obs}}}.
\end{equation}

\noindent The above equation can be rewritten as

\begin{equation}
\Omega_{\rm GW} = {\nu_{\rm{obs}}\over c^{3}\rho_{\rm c}}F_{\nu} = {4{\rm \pi}^{2}\over
3H^{2}_{0}}\nu_{\rm{obs}}^{2} h_{\rm BG}^{2}.
\end{equation}

In the next section we present the numerical results and discussions, which come mainly
from the equation for ${h_{\rm BG}}$.

\section{NUMERICAL RESULTS AND DISCUSSIONS}

In this section we present the numerical results of the putative GW background related to the NS phase transition.

The Eq. 15 is the main equation to calculate such a GW background, with the relevant ingredients given by Eqs. $ 16 - 24$. Basically, to integrate Eq. 15 one needs to specify the cosmological model, the star formation rate density, etc. We refer the reader to Section 4 to see in detail how the various relevant ingredients are taken into account.

More explicitly, to obtain the spectrum of the background using Eq. 15 one takes a given $\nu_{\rm obs}$ and perform integrations in the variables $m$ and $z$, where the integration intervals are those of the models present in Table 2.
This procedure is repeated for different $\nu_{\rm obs}$'s, whose values are taken from the frequency band also present in Table 2.

\begin{figure}
\includegraphics[width=0.75\textwidth]{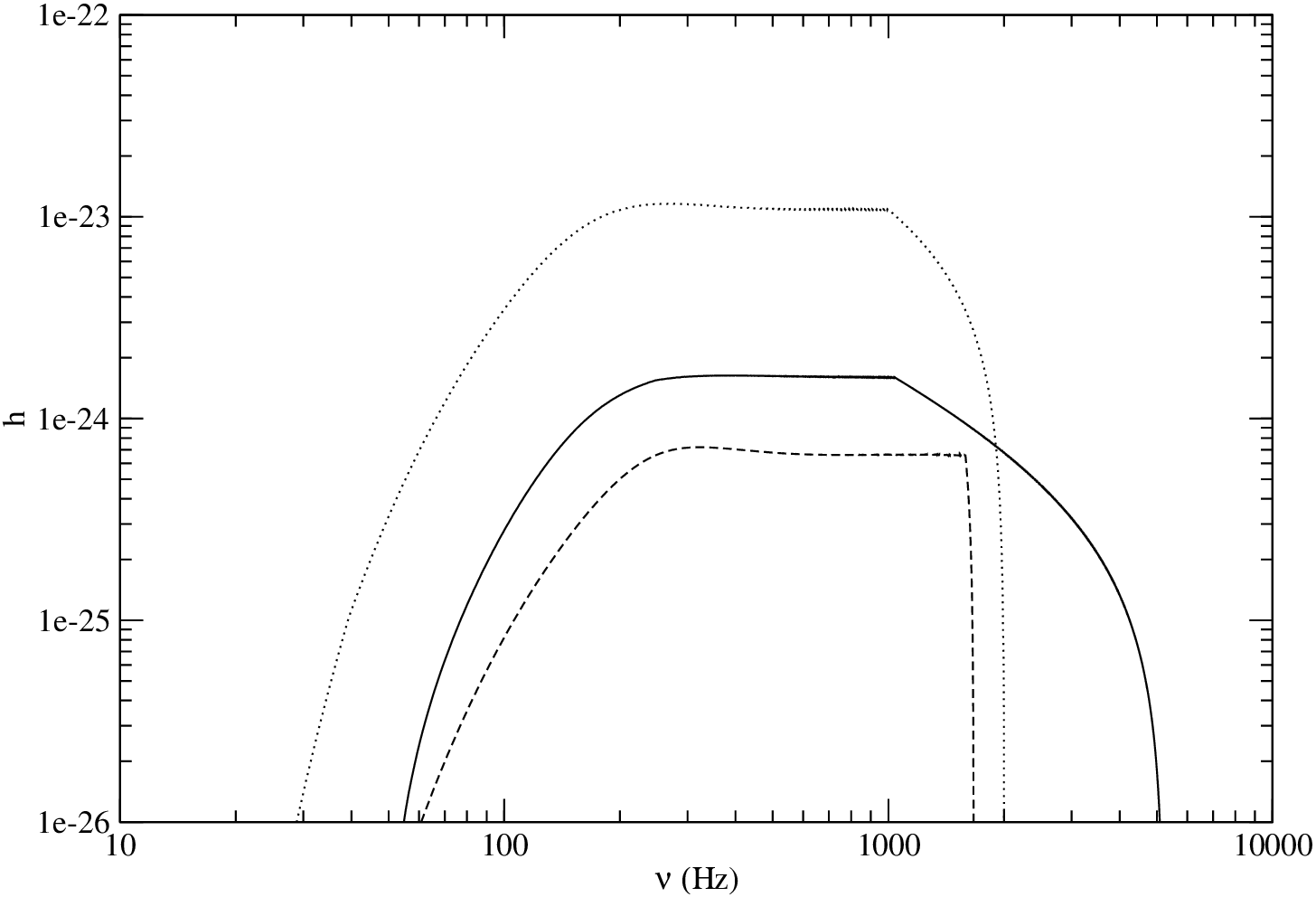}\vspace{1cm} \caption{\label{2} The
dimensionless
amplitude for the GW background of  stellar BH formation (solid line), the whole
ensemble of NSs
  undergoing phase transition (dotted line) and 1 \% of the NSs doing
  so (dashed line), for $0<z<20$.}
\end{figure}

In Figure \ref{2} we compare the background of GWs generated by NSs, which undergo phase
transition, with that by the BH formation, which undergo quasi-normal mode instability (see \cite{araujo2004} for details). In particular, for the NSs we consider two situations.

We consider in Figure 2 two versions of the model B (see Table 2): one with its original values, namely, $f_{\rm NS}^{pt} \sim 0.01$ and the other one, just as an example, with  $f_{\rm NS}^{pt} =1$. Obviously, even the most favorable values for nuclear and subnuclear matter coupling constants could not make $f_{\rm NS}^{pt}$ even tend to 1, but this model is included just as a matter of comparison, as it would represent an upper limit. In both cases we consider $E\sim 0.01M_\odot c^2$, which means that approximately only a tenth of the energy generated in phase transition goes to the pulsating mode. Note that we are considering an energy release in GW about ten times higher then \cite{sigl}. Later on we comment on how this choice modifies our conclusions concerning the putative detectability
of the background studied here.

Note that the background of GWs generated by the NSs would have an amplitude greater than
that generated by the BHs only if a considerable fraction of the NSs formed undergo phase
transition.

\begin{figure}
\includegraphics[width=0.75\textwidth]{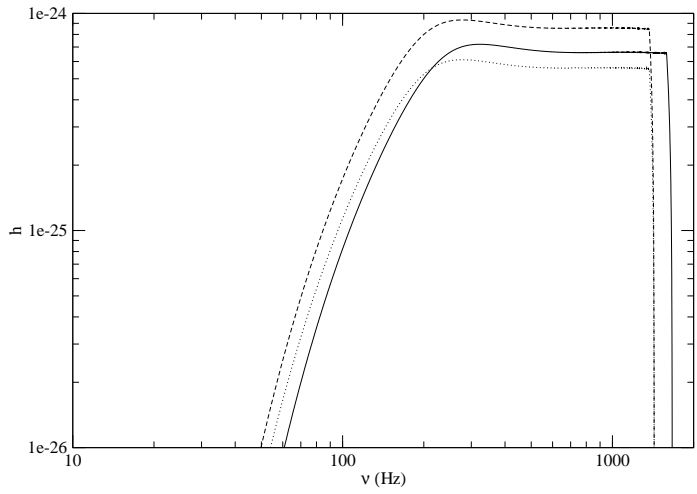}\vspace{1cm} \caption{\label{3} The
dimensionless
amplitude for the GW background
  of micro-collapsed NSs for $1.2<M<1.26M_\odot$ (solid line),
  $1.4<M<1.46M_\odot$ (dotted line) and a sharper NS distribution with
  $1.4<M<1.46M_\odot$ (dashed line). We consider in these models $f_{\rm NS}^{pt} = 0.01$
  and $E = 0.01M_{\odot}c^{2}$}
\end{figure}

In Figure \ref{3} we compare how different laws to calculate the NSs mass, the remnant
mass, for a Salpeter IMF, modify the spectrum of the background of GWs generated. Note that there is no significantly difference in the background for the three cases considered.

\begin{figure}
\includegraphics[width=0.75\textwidth]{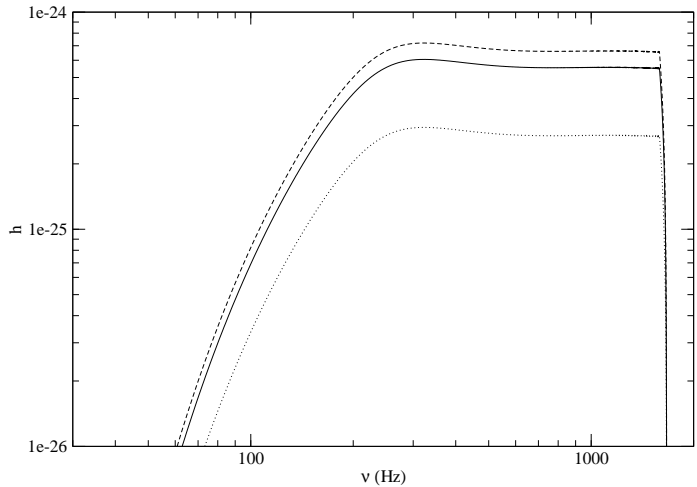}\vspace{1cm} \caption{\label{4} The
dimensionless
amplitude for the GW background of  micro-collapsed NSs for a IMF with $x=1.30$ (solid line),
$0.30$ (dotted line) and $1.85$ (dashed line). We consider in these models $f_{\rm NS}^{pt}
= 0.01$
  and $E = 0.01M_{\odot}c^{2}$.}
\end{figure}

We also investigated the role of the IMF in our results. In particular, besides the
standard IMF ($x=1.30$) we also consider two others, namely, $x=0.30$ and 1.85. This same
choice was considered by \cite{araujo2004} in their study on the background of GWs
generated by the formation of population III stellar BHs. As with for the BHs the amount
of NSs increases (decreases) for $x=0.30$ (1.85) with respect to $x=1.30$. In Figure \ref{4}
we compare these three cases. In this comparison, we consider model B.

\begin{figure}
\includegraphics[width=0.75\textwidth]{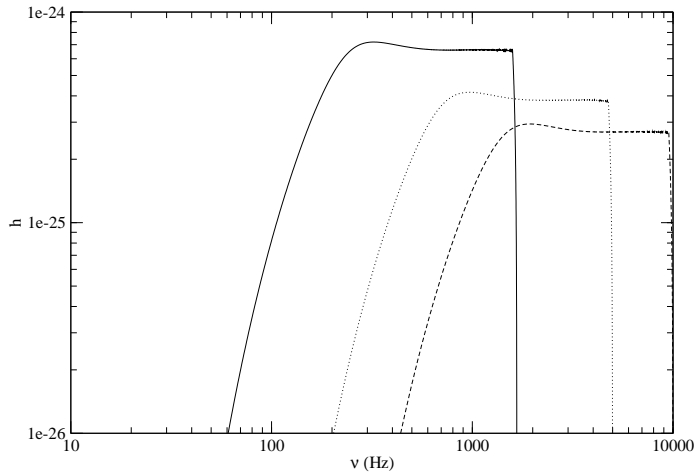}\vspace{1cm} \caption{\label{5} The
dimensionless amplitude for the GW background of micro-collapsed NSs considering the contributions of
different quasi-normal modes. The f-mode (solid line), the p-mode at $6kHz$ (dotted line)
and the w-mode at $12kHz$ (dashed line). We adopted model B (f-mode) with $f_{\rm NS}^{pt}
= 0.01$ and $E = 0.01M_{\odot}c^{2}$. For the p- and w- modes we also considered for comparison
these very two parameters.}
\end{figure}

We now compare model B, where we consider $f_{\rm NS}^{pt} \sim 0.01$ and $E \sim 0.01
\,M_{\odot}c^{2}$, with the spectra of backgrounds of GWs for two other pulsating modes,
considering these two very parameters. In particular, we present in Figure \ref{5} the
GW background of micro-collapsed NSs considering the contributions of the p-mode at 6 kHz
(dotted line) and the w-mode at 12 kHz (dashed line). As expected, see Figure \ref{5},
the higher the frequency is, the more the spectrum is shifted to the higher frequencies.
Concerning the amplitude, the spectra do not present considerable differences among them.

It is worth mentioning that we did not take into account the g-mode in our calculations
because the amount of energy put into this mode, which is related to density discontinuities
\cite{miniutti}, is much smaller then in the case of f- and p-modes and the frequencies
related to this mode lie in a lower range of the spectrum.

A relevant question is whether the background we study here is continuous or not. The duty
cycle indicates if the collective effect of the bursts of GWs generated during the collapse
of a progenitor star generates a continuous background. The duty cycle is defined as
follows:

\begin{equation}
DC = \int_{z_{\rm cf}}^{z_{\rm ci}} dR_{\rm NS}\bar{\Delta\tau_{\rm GW}}(1+z),
\end{equation}

\noindent where $\bar{\Delta\tau_{\rm GW}}$ is the average time duration of single bursts
at the emission, which amounts to $\sim 100\,{\rm ms}$ to the f-mode of NSs, as obtained
though the empirical equations of ref.\cite{benhar}. This value for the f-mode damping
time is quite different from that adopted by Sigl \cite{sigl}, who assumed a value
of $ 1/3\,{\rm ms}$.

Since the star formation rate could be high, a significant amount of GWs could be produced.
We also note that, independently of the primordial cloud mass and of the redshift of
collapse, star formation occurring at high redshift could produce high duty cycle values,
which lead us to conclude that the stochastic GW background could be continuous. For all
the models studied here the duty cycle is $> 1$.

In the next section we consider the detectability of the background of the GW background
that we propose exists.

\section{DETECTABILITY OF THE BACKGROUND OF GRAVITATIONAL WAVES}

The background predicted in the present study cannot be detected by single interferometric
detectors, such as VIRGO and LIGO (even by the forthcoming advanced ones). However, it is
possible to correlate the signal of two or more detectors to detect the background that we
propose exists. Christensen \cite{michel} was the first to show that this kind of signal can, in
principle, be detected by correlating the outputs of two different detectors. However, the
main requirement that must be fulfilled is that they must have independent noise. This
study was improved by \cite{christ} and by \cite{flanagan}. The reader should also refer to
the papers by \cite{allen1997} and \cite{allen1999} who also deal in detail with such an
issue.

To assess the detectability of a GW signal, one must evaluate the signal-to-noise ratio
(S/N), which for a pair of interferometers is given by (see, e.g.,
\cite{flanagan,allen1997})

\begin{equation}
\label{sn} {\rm (S/N)}^2=\left[\left(\frac{9 H_0^4}{50\pi^4} \right) T \int_0^\infty d\nu
\frac{\gamma^2(\nu)\Omega^2_{GW}(\nu) } {\nu^6 S_h^{(1)}(\nu) S_h^{(2)}(\nu)} \right]
\end{equation}

\noindent where $ S_h^{(i)}$ is the spectral noise density, $T$ is the integration time and
$\gamma(\nu)$ is the overlap reduction function, which depends on the relative positions
and orientations of the two interferometers. For the $\gamma(\nu)$ function we refer the
reader to \cite{flanagan}, who was the first to calculate a closed form for the LIGO
observatories. Flanagan \cite{flanagan} (see also \cite{allen1997}) showed that the best window for
detecting a signal is $0< \nu < 64$ Hz, where the overlap reduction function has the
greatest magnitude.

Here we consider, in particular, the LIGO interferometers: the Initial, the Enhanced and the Advanced ones.
For the Initial LIGO, the present configuration, the sensitivity curve has been taken from http://www.ligo.caltech.edu/~jzweizig/distribution/ LSC\_Data/. It is worth mentioning that the design-sensitivity
curve was almost attained by the S5 run. For the Enhanced LIGO, which is an upgrade of the Initial LIGO
and to be operative briefly, the sensitivity curve has been taken from http://www.ligo.caltech.edu /docs/T/T060156-01.pdf, in particular from its Fig. 12. The Enhanced LIGO is approximately a factor of two more sensitive than the Initial LIGO.
Finally, for the Advanced LIGO, designed to be approximately a factor of ten more sensitive than the Initial LIGO,
the sensitivity curve has been taken from http://lhocds.ligo-wa.caltech.edu:8000 /advligo/LSC\_Modeling?action=AttachFile\&do=get\&target=isccddFeb19.pdf, in particular from its Fig. 1.

We show in Table 3 the S/N for the models of Table 2 for the three different LIGO
generations and for different values of $f^{pt}_{NS}$.

\begin{table}
\caption {For the models of Table 2 with different values of $f^{pt}_{NS}$, we present the
S/N for pairs of Initial, Enhanced and Advanced LIGO observatories for one year of observation.
We consider in the calculations $E = 0.01 M_{\odot}c^{2}$.}
\begin{center}
\begin{tabular}{ccccc}
\hline & & &  S/N &   %\\ %\ns \ns \ns \ns %& &  \crule{3}
% \\ Model & $f^{pt}_{NS}$ & LIGO I & LIGO II & LIGO III \\
\\ Model & $f^{pt}_{NS}$ &Initial LIGO  & Enhanced LIGO  & Advanced LIGO  \\ \hline
A & 1.0  & $8.6\times 10^{-4}$  & $2.0\times 10^{-3}$ & $2.4\times 10^{-1}$  \\
B & 0.01 & $2.2\times 10^{-6}$  & $4.8\times 10^{-6}$ & $2.0\times 10^{-4}$  \\
C & 0.01 & $2.2\times 10^{-6}$  & $4.8\times 10^{-6}$ & $2.0\times 10^{-4}$  \\
D & 0.01 & $2.8\times 10^{-6}$  & $6.3\times 10^{-6}$ & $2.6\times 10^{-4}$  \\
E & 1.0  & $1.1\times 10^{-4}$  & $1.5\times 10^{-3}$ & $3.0\times 10^{-2}$  \\
F & 1.0  & $5.6\times 10^{-6}$  & $1.0\times 10^{-5}$ & $5.7\times 10^{-4}$  \\
G & 1.0  & $6.5\times 10^{-4}$  & $1.5\times 10^{-3}$ & $7.6\times 10^{-2}$  \\ \hline
\end{tabular}
\end{center}
\end{table}

As shown in Table 3, the signal-to-noise ratio for all models studied is lower than one,
even for an advanced LIGO. Therefore, contrary to the claim of \cite{sigl} such a putative
GW background would hardly be detected.

Using the same parameters of model B with $f_{\rm NS}^{pt} = 0.01$ and $E =
0.01M_{\odot}c^{2}$, we have also investigated the role of the variations in the IMF
modifying the exponent $x$. We have found, for the Advanced LIGO, a signal-to-noise ratio of
$3.4\times10^{-5}$, for $x=0.30$, and $1.5\times10^{-4}$, for $x=1.85$. As before, the
signal-to-noise ratios are lower than 1.

We still have to mention the results for different frequencies, which correspond to the
excitation of p- and w-modes. From our calculations, with the same parameters shown for
model B , i.e., with $f_{\rm NS}^{pt} = 0.01$ and $E = 0.01M_{\odot}c^{2}$, we have changed
the frequencies to $\nu_e=$ 3, 6 and 12 kHz, we obtain a signal-to-noise ratio for Advanced LIGO
equal to $2.8\times10^{-5}$, $7.8\times10^{-6}$ and $1.3\times10^{-8}$, respectively. These
results show that such background would not either be detected.

Note that the signal-to-noise ratio, for given IMF, $\alpha_{i}$, and integration time,
depends on $f_{\rm NS}^{pt}$ and $E$ as follows

\begin{equation}
{\rm (S/N)} \propto f_{\rm NS}^{pt} E \, ;
\end{equation}

\noindent and it also depends on the SFR density in a more complicated way, namely, through
an integral involving the redshift $z$. The higher the star formation rate, the higher the
signal-to noise ratio will be.

Just as a matter of comparison, even considering models with $f_{\rm NS}^{pt}=1$, a
signal-to-noise ratio significantly greater than one for the Advanced LIGO would be
possible either the SFR density would be much greater than that by Springel \& Hernquist,
or the energy generated in the phase transition were almost completely channeled to excite
the f- mode. Obviously, an optimist combination of the SFR density and the energy channeled
to the f-mode would also render the same.

Still concerning the energy release in GWs, the above equation shows explicitly how the detectability
is affected for lower values of $E$. For example, adopting an energy release as that by \cite{sigl} the
signal-to-noise ratios presented in Table 3 would be ten times lower.

\section{CONCLUSIONS}

We present here a study concerning the generation of GWs produced from a cosmological
population of NSs. These stars may undergo a phase transition if born close to the
transition density, suffering a micro-collapse and exciting quasi-normal modes.

This work differs from the previous one done by Sigl in two essential points. First, we have
calculated the NS star structure using realistic equation of state, instead of a polytropic model, obtained through a field-theoretical model which takes into account the microstructure of matter composed by the whole baryon octet, coupled to meson fields. We have also taken into account, into the quark sector, the standard MIT bag model. The transition is constructed via Gibbs criteria. We have calculated the amount of energy released by the transition and the rate of stars that undergo this microcollapse. Second, we have used another scenario for the cosmical star formation rate. We also have made a study considering various scenarios varying the set of parameters that gives the main characteristics of star formation masses, etc.

We show that a detectable background is possible only if the SFR density is much greater
than that predicted by Springel \& Hernquist or if the energy generated in the phase
transition is almost completely channeled to excite the f- mode. Obviously, a too
optimistic combination of these two possibilities could do the same.

It is worth mentioning that even planned detectors like DECIGO and BBO would not be sufficiently sensitive to detect the background considered here.

Finally, it is worth mentioning that Sigl claims a putative marginal detection if the majority of NSs
undergo a phase transition. In our study we obtain a similar conclusion for some of our models
if the above condition is fulfilled ($f_{\rm NS}^{pt}=1$) and for an integration time significantly
larger than one year.

One could ask why we obtained such a conclusion even adopting a release of GW energy larger than
that adopted by Sigl. One reason for that is related to the SFR densities we adopted, namely, in his
case a larger number of NSs is generated. As a matter of comparison, if you multiplied the SFR density
by a factor of ten the signal-to-noise ratio would be multiplied by this very factor. Other reasons
could be related to the way we calculate the NSs remnants (see Table 3) and the value Sigl adopted
for the damping time. A damping time of $\sim 100\,{\rm ms}$ instead of one of $ 1/3\,{\rm ms}$ could
modify his results significantly.

\section*{Acknowledgments}
GFM and JCNA would like to thank CNPq (grants 381682/2006-4 and 307424/2007-3, respectively) for financial support. The authors would like also to thank the referees for useful suggestions and criticisms. In particular, we would like to thank one of the referees who called our attention to the correct LIGO nomenclature as well as for kindly sent us the updated LIGO'S sensitivity curves.

\end{document}